\documentclass{llncs}

\usepackage{amsmath}
\usepackage{float}
\usepackage{enumerate}
\usepackage{url,epsfig}
\usepackage[all]{xy} 
\usepackage{tabularx}
\usepackage{multirow}
\usepackage{graphicx}
\usepackage[usenames,dvipsnames]{color}
\usepackage{caption}
\usepackage{float}
\usepackage{subfig}
\usepackage{mathtools}
\usepackage{bbm}
\usepackage[noend]{algpseudocode}
\usepackage{algorithm}
\usepackage{booktabs}
\usepackage{enumitem}

\begin{document}

\title{Improving Entity Retrieval on Structured Data}

\author{Besnik Fetahu, Ujwal Gadiraju and Stefan Dietze}
\institute{ 
	L3S Research Center, Leibniz Universit\"at Hannover, Germany$\>\>$ 
	\\ \{fetahu, gadiraju, dietze\}@L3S.de \linebreak[1]
}
\maketitle

\begin{abstract}
The increasing amount of data on the Web, in particular of Linked Data, has led to a diverse landscape of datasets, which make entity retrieval a challenging task. Explicit cross-dataset links, for instance to indicate co-references or related entities can significantly improve entity retrieval. However, only a small fraction of entities are interlinked through explicit statements. In this paper, we propose a two-fold entity retrieval approach. In a first, offline preprocessing step, we cluster entities based on the \emph{x--means} and \emph{spectral} clustering algorithms. In the second step, we propose an optimized retrieval model which takes advantage of our precomputed clusters. For a given set of entities retrieved by the BM25F retrieval approach and a given user query, we further expand the result set with relevant entities by considering features of the queries, entities and the precomputed clusters. Finally, we re-rank the expanded result set with respect to the relevance to the query. We perform a thorough experimental evaluation on the Billions Triple Challenge (BTC12) dataset. The proposed approach shows significant improvements compared to the baseline and state of the art approaches. 
\end{abstract}

\section{Introduction}\label{sec:introduction}

The emergence of the Web of Data, particularly supported through W3C standards such as RDF and the Linked Data principles~\cite{DBLP:journals/ijswis/BizerHB09}, has led to a wide range of semi-structured RDF data being available on the Web. Data is spread across datasets, complemented through a growing amount of entities as part of structured annotations of Web documents, using RDFa or Microformats. Recent studies have shown that approximately 26\% of pages already contain structured annotations~\cite{meusel@iswc2014}.

Web data forms a highly heterogeneous knowledge-graph spanning an estimated 100 billion triples~\cite{DBLP:conf/www/PoundMZ10}, with a wide variety of languages, schemas, domains and topics~\cite{DBLP:conf/esws/GueretGSL12}. Even though a large number of entities and concepts are highly overlapping, that is they represent the same or related concepts, explicit links are still limited and often concentrated within large established knowledge graphs, like DBpedia~\cite{DBLP:conf/semweb/AuerBKLCI07}.

The entity-centric nature of the Web of data has led to a shift towards tasks related to entity and object retrieval~\cite{DBLP:conf/semweb/BlancoCMT13,DBLP:conf/sigir/TononDC12} or entity-driven text summarization~\cite{DBLP:conf/sigir/DemartiniMBZ10}. Major search engine providers such as Google and Yahoo! already exploit such data to facilitate semantic search using knowledge graphs, or as part of similar efforts such as the \textit{EntityCube-Renlifang} project at Microsoft Research~\cite{DBLP:conf/www/NieMSWM07}. In such scenarios, data is aggregated from a range of sources calling for efficient means to search and retrieve entities in large data graphs. 
Specifically, \emph{entity retrieval} (also known as Ad-Hoc Object retrieval)~\cite{DBLP:conf/www/PoundMZ10,DBLP:conf/sigir/TononDC12} aims at retrieving relevant entities given a user query. The result is a ranked list of entities~\cite{DBLP:conf/semweb/BlancoCMT13}. By simply applying standard keyword search algorithms, like the BM25F, promising results can be achieved. A common practice is to construct indexes over the textual descriptions (\emph{literals}) of entities. 

In most cases, queries are entity centric. However, there are a large number of queries that are also topic-based, e.g. `\texttt{U.S. Presidents}'. Therefore, approaches like~\cite{DBLP:conf/sigir/TononDC12} have proposed retrieval techniques that make use of the explicit links between entities in the WoD for results or query expansion. For instance, following \texttt{owl:sameAs} or \texttt{rdfs:seeAlso} predicates from \texttt{dbp:Barack\_Obama}, one can retrieve co-references or highly related entities. However, considering the size of the WoD such statements are very sparse (see Figure~\ref{fig:related_types}).

In this work, we propose a method for improving entity retrieval results in two aspects. We improve the task by \emph{expanding} and \emph{re-ranking} the result set from a baseline retrieval model (BM25F). Sparsity of explicit links is addressed through clustering of entities based on their similarity, using a combination of lexical and structural features. Consequently, we expand the result set with additional entities from the \emph{cluster space} (clusters with which the baseline entities are associated), retrieved from the baseline. 

For the expanded result set, there is a need for re-ranking. The re-ranking considers the similarity of entities to the user query, and their relevance likelihood based on the corresponding entity type, defined as \emph{query type affinity}. We empirically model the query type affinity between the entity type in a query (e.g. \emph{`Barack Obama'} \texttt{isA Person}) and the entity types in the result set (see Section \ref{subsec:motivation_approach}).

In terms of \emph{scalability} and \emph{efficiency}, the clustering process is carried out offline, where we \emph{bucket} entities of particular types together before clustering. This improves the efficiency by reducing the run-time of the clustering algorithms (Section~\ref{subsec:bucketing_clustering} and \ref{subsec:discussion}). The entity retrieval, expansion and re-ranking on the other hand are performed online and the computational overhead is negligible (Section~\ref{sec:retrieval} and \ref{subsec:discussion}).

Our experimental evaluation is carried out on the BTC12 dataset~\cite{btc-2012}, and using the SemSearch\footnote{\url{http://km.aifb.kit.edu/ws/semsearch10/}} query dataset. The individual steps in our approach are evaluated through a reliable crowdsourced evaluation approach. The results show that the proposed approach outperforms existing basslines for the entity retrieval task.

The main contributions of our work are as follows: (a) an entity retrieval model combining keyword search and entity clustering, and (b) an entity ranking model considering the query type affinity w.r.t the set of relevant entity types.

\section{Related Work}\label{sec:relatedwork}
A large portion of queries issued in Web search engines target entities or contain semantic resources (such as types, relations and attributes) \cite{DBLP:conf/www/PoundMZ10} as a primary intent. Consequently, the identification of entity-centric queries has become of particular concern for commercial search engines serving as a means to narrow the search space and to provide contextual query results \cite{Mika:2009:ISG:1693684.1693713}. Thus, the traditional task of Ad-hoc Document Retrieval (ADR) \cite{Manning:2008:IIR:1394399} is moving towards an entity retrieval task \cite{DBLP:conf/www/PoundMZ10}. Hence, instead of top--$k$ document retrieval that match a keyword query, the task and therefore the results are increasingly becoming entity-centric. 

Following this direction, Tonon et al. \cite{DBLP:conf/sigir/TononDC12} proposed a hybrid approach based on query expansion and relevance feedback techniques on top of the BM25 ranking function to build an entity retrieval framework. In contrast to this work, we use the state-of-the-art BM25F \cite{Blanco:2011:EEE:2063016.2063023,RobertsonBM25F} to assign varying degrees of importance to different parts of a document. Further, through an offline pre-processing step we are able to infer links between similar entities for the retrieval process. This is particularly important when considering datasets that have less links between entities, a significant feature of the work by Tonon et al \cite{DBLP:conf/sigir/TononDC12}. Another advantage of adopting BM25F is penalising documents/entities, consisting of long textual literals, in the final ranking \cite{Lv:2011:DVL:2009916.2010070}. Sindice~\cite{DBLP:journals/ijmso/OrenDCCST08} is another approach focusing on indexing RDF documents. It supports data discovery and integration by taking advantage of DBpedia entities as a source to actively index resources. The process performed by Sindice plays a key role in centralising disparate data sources on the Web. The adoption of entities and foremost entity types (topics) is also supported by \cite{DBLP:conf/semweb/BlancoCMT13} in the recommendation of entities in Web search. Our approach can benefit Sindice by indexing documents following a topic-based fashion. 

Zhiltsov and Agichtein~\cite{Zhiltsov:2013:IES:2541154.2507868} propose a learning to rank approach, where they model the relations between entities through a various set of features, such as language models and other query related features (e.g query length). Finally, through tensor matrix factorisation they find latent similarities between entities, later used in their learning to rank model. One major disadvantage of this approach is that it is supervised, hence, unlikely to perform reasonably well on ad-hoc entity search tasks.

\section{Approach and Overview}\label{sec:motivationbackground}
In this section, we motivate and define our work in the context of the addressed challenge, and provide an overview of our approach.

\subsection{Preliminaries}\label{subsec:preliminaries}
The \textit{entity retrieval (ER) task}, also known as ad-hoc object retrieval, is concerned with retrieving a top--$k$ ranked set of entities from dataset for a given a user query $q$. User queries are typically entity centric. A \emph{dataset} in our case is a set of triples $\langle s, p, o \rangle$, where $s$ is the \emph{subject} (the URI of an entity), $p$ is the \emph{predicate}, and $o$ is the \textit{object} (a URI or a literal). An \emph{entity} profile of $e$ is the set of triples sharing the same subject URI $s$. The \emph{type} of an entity is determined by the object of the triple $t_e=\langle s,$ \texttt{rdf:type}$, o \rangle$. 
Additionally, we define the \emph{query type} $t_q$, corresponding to the entity type in $q$, e.g. \emph{`Barack Obama'}, hence $t_q $\texttt{ hasType Person}. 

\subsection{Motivation: Result Set Expansion and Query Affinity in Entity Retrieval} \label{subsec:motivation_approach}

Recent studies \cite{DBLP:conf/sigir/TononDC12} have shown that \emph{explicit similarity statements}, which indicate some form of similarity or equivalence between entities, for instance through predicates such as $\texttt{owl:sameAs}$, are useful for improving entity retrieval results as retrieved through approaches like BM25F, i.e. improving significantly on standard precision/recall metrics. However, such explicit similarity statements usually are sparse and often focused towards a few well established datasets like DBpedia, Freebase etc. One main reason is that these datasets represent known, and well structured graphs, which show a comparably high proportion of such dedicated similarity statements, in turn linking similar entities within and beyond their original namespace.

\begin{figure}[ht!]
\centering
    \includegraphics[width=\textwidth]{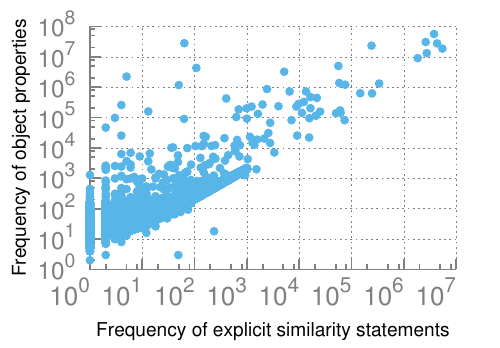}
    \caption{Number of explicit similarity statements in contrast to the frequency of object property statements overall, shown for all data graphs.}
    \label{fig:related_types}
\end{figure}

In Figure~\ref{fig:related_types} we show the total amount of explicit similarity statements (on the x--axis) that interlink entities in the BTC12 dataset. Referring to \cite{DBLP:conf/sigir/TononDC12}, here we specifically consider triples of the form $\langle e, p, e'\rangle$ where the predicate $p\in\{$\texttt{owl:sameAs}, \texttt{skos:related}, \texttt{dbp:wikiPageExternalLink}, \texttt{dbp:wikiPageDisambiguates}, \texttt{dbp:synonym}$\}$. These are plotted against the total number of \emph{object properties} (y--axis), where each point in the plot represents a graph in the BTC12 collection. From the figure, it is obvious that the number of explicit similarity statements is very sparse, considering the size of the dataset.

\begin{figure}[ht!]
	\includegraphics[width=\textwidth]{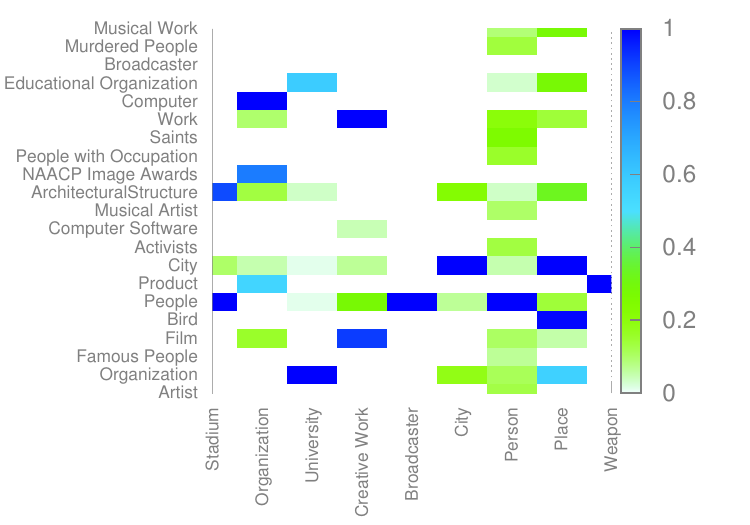}
	\caption{ Query type affinity shows the query type and the corresponding entity types from the retrieved and relevant entities.}
	\label{fig:query_affinity}
\end{figure}

Nonetheless, missing links between entities can be partially remedied by computing their pair-wise similarity, thereby complementing statements like \texttt{owl:sameAs} or \texttt{skos:related}. Given the semi-structured nature of RDF data, graph-based and lexical features can be exploited for similarity computation. Particularly, lexical features derived from literals provided by predicates such as \textit{rdfs:label} or \textit{rdfs:description} are prevalent in LOD. Our analysis on the BTC12 dataset reveals that a large portion of entities (around 90\%) have an average literal length of 50 characters.

Furthermore, while the query type usually is not considered in state of the art ER methods, we investigated its correlation with the corresponding entity types from the query result set. We refer to a ground truth\footnote{\url{http://km.aifb.uni-karlsruhe.de/ws/semsearch10/Files/assess}} using the BTC10 dataset. We focus only on relevant entities for $q$. We analyze the \emph{query type affinity} of the result sets by assessing the likelihood of an entity in the results to be of the same type as the query type. Figure~\ref{fig:query_affinity} shows the query type affinity. On the x-axis we show the query type, whereas on the y-axis the corresponding relevant entity types are shown. Figure~\ref{fig:query_affinity} shows that most queries have high affinity with a specific entity type, with the difference being the query type \texttt{Person}, where relevant entities have a wider range of types.

Our work exploits such \emph{query type affinity} to improve the ranking of entities for a query $q$ (see Section~\ref{sec:retrieval}). Based on these observations, we argue that (a) \emph{entity clustering} can remedy the lack of existing linking statements and (b) entity \emph{re-ranking} considering the \emph{query type affinity} are likely to improve the entity retrieval task.

\subsection{Approach Overview} 
\begin{figure}[ht]
\centering
\includegraphics[width=0.95\textwidth]{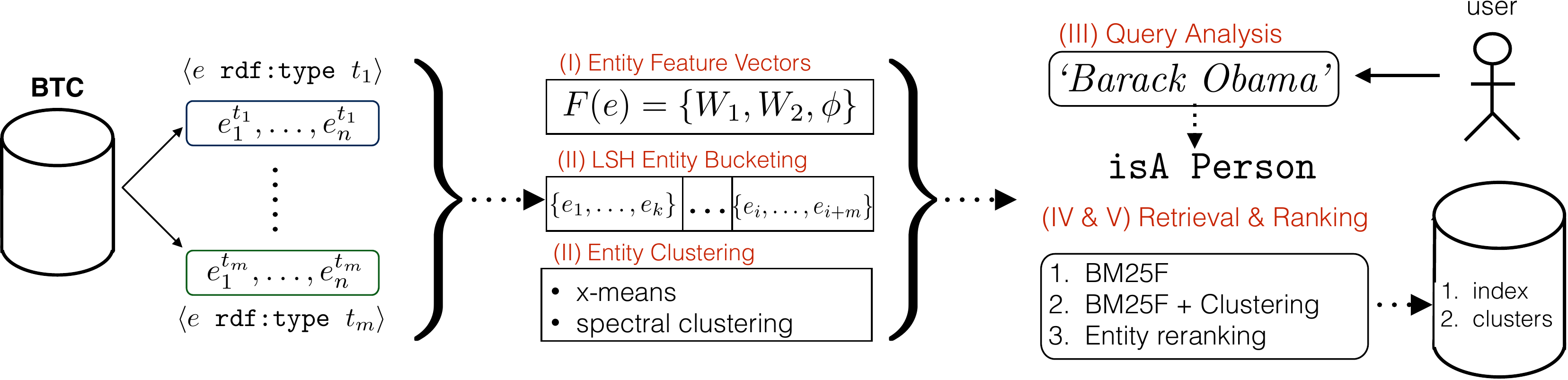}
\caption{\small{Overview of the \emph{entity retrieval} approach.}}
\label{fig:workflow}
\end{figure}

In this work we propose a novel approach for the \emph{entity retrieval} task which builds on the observations described earlier. Figure~\ref{fig:workflow} shows an overview of the proposed approach. The individual steps are outlined below and described in detail in Section~\ref{sec:preprocessing} and \ref{sec:retrieval}. We distinguish between two main steps: (I) \emph{offline pre-processing}, including step I.a and I.b in the following overview, and (II) \emph{online entity retrieval}, covered by steps II.a to II.c.

\textbf{I.a Entity Feature Vectors:} We construct the entity feature vector as follows: $F(e)=\{\mathbf{W}_1(e),\mathbf{W}_2(e), \phi\}$, where $\mathbf{W}_1(e)$ and $\mathbf{W}_2(e)$ represent the \emph{unigrams} and \emph{bigrams} extracted from literals of $e$, and $\phi$ represents the structural features. 

\textbf{I.b Entity Bucketing \& Clustering:} is used to compute implicit relationships between entities emerging from their feature vectors. For the sake of efficiency, before we proceed with entity clustering, we exploit the locality-sensitive hashing (LSH) algorithm for bucketing. 

\textbf{II.a Query Analysis:} As part of the retrieval task, we initially analyse the given user queries $q$. From the query terms, which typically represent named entities, we determine the type of the named entity, e.g. `\texttt{Location}' in order to support the query type affinity-based reranking at a later stage.

\textbf{II.b Entity Retrieval:} In the retrieval process, we rely on a combination of standard IR approaches, like BM25F and further expand the result set with entities showing a high similarity according to the computed clusters.

\textbf{II.c Entity Ranking.} In the final step, we rank the expanded entity result set for $q$, taking into account similarity to the query and the modelled query type affinity.

\section{Data Pre-processing and Entity Clustering}\label{sec:preprocessing}

In this section, we describe the offline pre-processing to cluster entities and remedy the sparsity of explicit entity links.

\subsection{Entity Feature Vectors} 
Entity similarity is measured based on a set of structural and lexical features, denoted by the \emph{entity feature vector} $F(e)$. The features for clustering are described below.

\textbf{Lexical Features:} We consider a weighted set of \emph{unigrams} and \emph{bigrams} for an entity $e$, by extracting all textual literals used to describe $e$ denoted as $\mathbf{W}_1(e)$ and $\mathbf{W}_2(e)$. The weights are computed using the standard \emph{tf--idf} metric. Lexical features represent core features when considering the entity retrieval task, more so for the clustering process. A high lexical similarity between an entity pair is a good indicator for expanding the result set from the corresponding cluster space.

\textbf{Structural Features:} The feature set $\phi(e)$ considers the set of all object properties that describe $e$. The range of values for the structural features is $\phi(o,e) \rightarrow [0,1]$, i.e., to indicate if a object value is present in $e$. 
\textbf{Feature Space:} To reduce the feature space, we filter out items from the lexical and structural features that occur with low frequency across entities and presumably, have a very low impact on the clustering process due to their scanty occurrence.

\subsection{Entity Bucketing \& Clustering}\label{subsec:bucketing_clustering}

\textbf{Entity Bucketing.} In this step we \emph{bucket} entities of a given \emph{entity type} by computing their \emph{MinHash} signature, which is used thereafter by the LSH algorithm~\cite{rajaraman2011mining}. This step is necessary as the number of entities is very large. In this way we reduce the number of pair-wise comparisons for the entity clustering, and limit it to only the set of entities within a bucket. Depending on the \emph{clustering algorithm}, the impact of bucketing on the clustering scalability varies. Since the LSH algorithm itself has linear complexity, bucketing entities presents a scalable approach considering the size of datasets in our experimental evaluation. A detailed analysis is presented in Section~\ref{sec:evaluation}. 

\textbf{Entity Clustering.} Based on the computed feature vectors, we perform entity clustering for the individual entity types and the computed LSH buckets. Taking into account scalability aspects of such a clustering process we consider mainly two clustering approaches: (i) \emph{X--means} and (ii) \emph{Spectral Clustering}. In both approaches we use Euclidean distance as the similarity metric. The dimensions of the Euclidean distance are the feature items in $F(\cdot)$. The similarity metric is formally defined in Equation~\ref{eq:entity_sim}.
\begin{equation}\label{eq:entity_sim}\small
	d(e, e') = \sqrt{\sum\left(\mathbf{F}(e) - \mathbf{F}(e')\right)^2}
\end{equation}
where the sum aggregates over the union of feature items from $\mathbf{F}(e), \mathbf{F}(e')$. The outcome of this process is a set of clusters $\mathbf{C}=\{C_1,\ldots,C_n\}$. The clustering process represents a core part of our approach from which we expand the entity results set for a given query, beyond the entities that are retrieved by a baseline as a starting point. The way the clusters are computed has an impact on the entity retrieval task, thus we present a thorough evaluation of cluster configurations in Section~\ref{subsec:clustering_eval}.

\textbf{X--means} 
To cluster entities bucketed together through the LSH algorithm and of specific entity types, we adopt an extended version of \emph{k-means} clustering, presented by Pelleg et al. which estimates the number of clusters efficiently \cite{pelleg2000x}. \emph{X--means} overcomes two major drawbacks of the standard \emph{k-means} clustering algorithm; (i) computational scalability, and (ii) the requirement to provide the number of clusters \emph{k} beforehand. It extends the \emph{k--means} algorithm, such that a user only specifies a range [$K_{min}$, $K_{max}$] in which the number of clusters, \emph{K}, may reasonably lie in. The bounds for $K$ in our case are set to $[2,50]$ clusters. 

\textbf{Spectral Clustering} 
In order to proceed with the \emph{spectral clustering} process, we first construct the adjacency matrix $\mathbf{A}$. The adjacency matrix corresponds to the similarity between entity pairs $d(e,e')$ of a given entity type and bucket. Next, from $\mathbf{A}$ we compute the unnormalised graph Laplacian~\cite{DBLP:journals/sac/Luxburg07} as defined in Equation~\ref{eq:graph_laplacian}:
\begin{equation}\label{eq:graph_laplacian}\small
\mathbf{L} = diag(\mathbf{A}) - \mathbf{A}
\end{equation} 
where, $diag(\mathbf{A})$ corresponds to the diagonal matrix, i.e., $diag(\mathbf{A})_{i,i}=\mathbf{A}_{i,j}$ for $i=j$.

From matrix $\mathbf{L}$ we are particularly interested in specific properties, which we use for \emph{clustering} and which are extracted from the \emph{eigenvectors} and \emph{eigenvalues} by performing a singular value decomposition on $\mathbf{L}$. The eigenvectors correspond to a square matrix $n\times n$, where each row represents the projected entity into a $n$-dimensional space. Eigenvectors are later used to cluster entities using standard \emph{k--means} algorithm. 

However, an important aspect that has impact on the clustering accuracy, is the number of dimensions considered for the \emph{k--means} and the $k$ itself. We adopt a heuristic proposed in~\cite{DBLP:journals/sac/Luxburg07}. The number of dimensions that are used in the clustering step corresponds to the first spike in the eigenvalue distribution. In addition, this heuristic is also used to determine the number $k$ for the clustering step.

\section{Entity Retrieval - Expansion and Reranking}\label{sec:retrieval}
In this section, we describe the online process of entity retrieval, including the process of \emph{expansion} and \emph{re-ranking} of the query result set.

\subsection{Query-biased Results Expansion}\label{subsec:entity_similarity}
Having obtained an initial result set $E_{b}=\{e_1,\ldots,e_k\}$ through a state of the art ER method (BM25f), the next step deals with expanding the result set for a given user query. From entities in $E_b$, we extract their corresponding set of clusters $\mathbf{C}$ as computed in the pre-processing stage. The result set is expanded with entities belonging to the clusters in $\mathbf{C}$. We denote the entities extracted from the clusters with $E_c$.

There are several precautions that need to be taken into account in this step. We define two threshold parameters for expanding the result set. The first parameter, \emph{cluster size}, defines a threshold with respect to the number of entities belonging to a cluster. If the number is above a specific threshold, we do not take into account entities from that cluster. The underlying rationale is that clusters with a large number of entities tend to be generic and less homogeneous, i.e. they tend to be a weak indicator of similarity. The second parameter deals with the \emph{number of entities} with which we expand the result set for a given entity cluster. The entities are considered based on their distance to the entity $e_b$. We experimentally validate the two parameters in Section~\ref{sec:evaluation}.

The \emph{fit} of expanded entities $e_c\in E_c$ concerns their similarity to query $q$ and the similarity to $e_b$, which serves as the starting point for the expansion of $e_c$. We measure the \emph{query-biased} entity similarity in Equation~\ref{eq:linkscore}, where the first component of the equation measures the \emph{string} distance of $e_c$ to $q$, that is $\varphi(q, e_c)$. Furthermore, this is done relative to entity $e_b$, such that if the $e_b$ is more similar to $q$, $\varphi(q,e_b) < \varphi(q,e_c)$ the similarity score will be increased, hence, the expanded entity $e_c$ will be penalized later on in the ranking (note that we measure distance, therefore, the lower the $sim(q,e)$ score the more similar an entity is to $q$). 

The second component represents the actual distance score $d(e_b, e_c)$.
\begin{equation}\label{eq:linkscore}\small
sim(q,e_c) = \lambda\frac{\varphi(q,e_c)}{\varphi(q,e_b)}  + (1-\lambda)d(e_{b}, e_{c})
\end{equation}
We set the parameter $\lambda=0.5$, such that entities are scored equally with respect to their match to query $q$ and the distance between entities, based on our baseline approach. The main outcome of this step is to identify possibly relevant entities that have been missed by the scoring function of BM25F. Such entities could be suggested as relevant from the extensive clustering approaches that consider the structural and lexical similarity.

\subsection{Query Analysis for Re-ranking}\label{subsec:query_analysis}
Following the motivation example in Figure~\ref{fig:query_affinity}, an important factor on the re-ranking of the result set is the \emph{query type affinity}. It models the relevance likelihood of a given entity type $t_e$ for a specific query type $t_q$. We give priority to entities that are most likely to be relevant to the the given query type $t_q$ and are least likely to be relevant for other query types $t_q'$. The probability distribution is modeled empirically based on a previous dataset, BTC10. The score $\gamma$, we assign to any entity coming from the expanded result set is computed as in Equation~\ref{eq:type_affinity}. 
\begin{equation}\label{eq:type_affinity}\small
\gamma(t_e, t_q) = \frac{p(t_e | t_q)}{\sum\limits_{t_q'\neq t_q} \left(1 - p(t_e | t_q')\right)}
\end{equation}

An additional factor we use in the re-ranking process is the \emph{context score}. To better understand the query intent, we decompose a query \emph{q} into its \emph{named entities} and additional \emph{contextual terms}. An example is the query $q=\{\text{\emph{`harry potter movie'}}\}$ from our query set, in which case the contextual terms would be `\emph{movie}' and the named entity `\emph{Harry Potter}' respectively. In case of ambiguous queries, the contextual terms can further help to determine the query intent. The \textit{context score} (see Equation~\ref{eq:context}) indicates the relevance of entity $e$ to the contextual terms $Cx$ of the query $q$. For entities with a high number of textual literals, we focus on the main literals like \emph{labels, name} etc.  
\begin{equation}\label{eq:context}\small
context(q,e) = \frac{1}{|Cx|}\sum\limits_{c_x \in Cx}{\mathbbm{1}_{e\text{\texttt{ has }} c_x}}
\end{equation}

\subsection{Top--$k$ Ranking Model}
The final step in our entity retrieval approach, re-ranks the expanded entity result set for a query $q$. The result set is the union of entities $\mathbf{E}=\mathbf{E}_b\cup\mathbf{E}_c$. In the case of entities retrieved through the baseline approach $e\in\mathbf{E}_b$, we simply re-use the original score, but normalize the values between $[0,1]$. For entities from $E_c$ we normalize the similarity score relative to the rank of entity $e_b$ (the position of $e_b$ in the result set) which was used to suggest $e_c$. This boosts entities which are the result of expanding top-ranked entities. 
\begin{equation}\small
rank\_score(e) = 
\begin{cases}
\frac{sim(q,e)}{rank(e_b)} & \text{ if } e \in E_c\\
bm25f(q,e) & \text{ otherwise }
\end{cases}
\end{equation}

The final ranking score $\alpha(e,t_q)$, for entity $e$ and query type $t_q$ assigns higher rank score in case the entity has high similarity with $q$ and its type has high relevance likelihood of being relevant for query type $t_q$. Finally, depending on the query set, in case $q$ contains contextual terms we can add $context(q,e)$ by controlling the weight of $\lambda$ (in this case $\lambda=0.5$).
\begin{align}\label{eq:rank}\small
\alpha(e,t_q) = \lambda \left(rank\_score(e) * \gamma(t_e, t_q)\right) + (1-\lambda) * context(q,e)
\end{align}
The score $\alpha$ is computed for all entities in $\mathbf{E}$. In this way based on observations of similar cases in previous datasets, like the BTC10 we are able to rank higher entities of certain types for specific queries.

\section{Experimental Setup}\label{sec:experimentalsetup}

Here we describe our experimental setup, specifically the datasets, baselines and the ground truth. The setup and evaluation data are available for download\footnote{\url{http://l3s.de/~fetahu/iswc2015/}}.

\subsection{Evaluation Data}
\textbf{Dataset.} In our experimental setup we use the BTC12 dataset~\cite{btc-2012}. It represents one of the largest periodic crawls of Linked Data, also containing well-known knowledge bases like Freebase and DBpedia. The overall statistics of the data are: (i) 1.4 billion triples, (ii) 107,967 graphs, (iii) 3,321 entity types, and (iv) 454 million entities.

\textbf{Entity Clusters.} The statistics for the generated clusters are as follows: the average number of entities fed into the \emph{LSH} bucketing algorithm is 77,485, whereas the average number of entities fed into \emph{x--means} and \emph{spectral} is 400. The number of generated entity buckets by LSH is 20,2009, while the number of clusters for \emph{x--means} and \emph{spectral} is 13 and 38, with an average of 10 and 20 entities per cluster respectively.

\textbf{Query Dataset.} To evaluate our retrieval approach we use the \emph{SemSearch}\footnote{\url{http://km.aifb.kit.edu/ws/semsearch10/}} query set from 2010 with 92 queries. The SemSearch query set is a standard collection for evaluating entity retrieval tasks.

\subsection{Baseline and State of the Art}

\textbf{Baseline.} We distinguish between two cases for the original BM25F baseline: (i) $\mathbf{B_{t}}$ and (ii) $\mathbf{B_{b}}$. In the first case, we use the \emph{title} or \emph{label} of an entity as a query field, whereas in the second case we use the full \emph{body} of an entity (consisting of all textual literals). The scoring of the fields is performed similar as in \cite{Blanco:2011:EEE:2063016.2063023}.

\textbf{State of the art.} We consider the approach proposed in~\cite{DBLP:conf/sigir/TononDC12} as the state-of-the-art. Similar to their experimental setup, we analyze two cases: (i) $\mathbf{S1}$ and (ii) $\mathbf{S2}$. $\mathbf{S1}$ expands the entity set from the baseline approach with directly connected entities, and $\mathbf{S2}$ expands with entities up to the second hop. For further details we refer the reader to \cite{DBLP:conf/sigir/TononDC12}. In our experiments, we found that the $\mathbf{S2}$ did not result in any significant change in performance when compared to $\mathbf{S1}$, and we therefore do not report further on $\mathbf{S2}$.

\textbf{Our approaches.} We analyze two entity retrieval techniques from our approach. The first is based on the \emph{x--means} clustering approach, which we denote by $\mathbf{XM}$. The second technique is based on \emph{spectral} clustering and is denoted by $\mathbf{SP}$. In both cases, we only expand the result set with entities coming from clusters with a total of ten entities associated with a cluster (see Section~\ref{subsec:entity_similarity}), and finally add only the most relevant entity based on the $sim(q,e_c)$ score. 

\textbf{BTC indexes.} For the baseline, we generate a Lucene index, where we index entity profiles on two fields \texttt{title} and \texttt{body} (consisting of all the textual literals of an entity). The second index is an RDF index over the BTC dataset with support for SPARQL queries, for which we use the RDF3X tool~\cite{Neumann:2008:RRE:1453856.1453927}. The first index is used for the baseline approach, while the second for the state of the art approach.

\subsection{Ground Truth for Evaluation of Entity Retrieval}
\label{ss:q}

For each query in the \emph{SemSearch2010} query set, we first establish the ground truth through crowdsourcing. Crowdsourced evaluation campaigns for the task of ad-hoc object retrieval have been shown to be reliable \cite{DBLP:conf/sigir/BlancoHHMPTT11,Halpin10evaluatingadhoc}. For each of the 92 queries, we pool the top 50 entities retrieved by the various methods, resulting in the top-k pooled entities corresponding to the query. By doing so we generate 4,600 query-entity pairs. 

We deploy atomic tasks in order to acquire relevance labels from the crowd for each query-entity pair. We follow the key prescriptions for task design and deployment that emerged from the work of Blanco et al. \cite{DBLP:conf/sigir/BlancoHHMPTT11} to build a ground truth. Workers are asked to assess the relevance of each retrieved entity to the corresponding query on a 5-point Likert-type scale\footnote{\textit{1:Not Relevant}, \textit{2:Slightly Relevant}, \textit{3:Moderately Relevant}, \textit{4:Fairly Relevant} and \textit{5:Highly Relevant}}. 

We collect 5 judgements from different workers for each pair to ensure reliable relevance assessments and discernible agreement between workers. This results in a total of 23,000 judgements. The final relevance of an entity is considered to be the aggregated relevance score over the 5 judgements. We assess and compare the performance of the different methods by relying on the ground truth thus generated (see Section 7).

\subsection{Evaluation Metrics}

Evaluation metrics assess the clustering accuracy and the retrieval performance.

\textbf{Cluster Accuracy.} As an initial evaluation, we assess the quality of our clusters. From a set of entities belonging to the same cluster, the accuracy is measured as the ratio of entities that \emph{belong together} over the total number of entities in a cluster, where assessments are obtained through crowdsourcing (see Section \ref{sec:evaluation}).

\textbf{Precision.} P@k measures the precision at rank $k$, in our case $k=\{1,\ldots,10\}$. It is measured as the ratio of retrieved and relevant entities up to rank $k$ over the total number of entities retrieved up to rank $k$.

\textbf{Recall.} R@k is measured as the ratio of retrieved and relevant entities up to rank $k$ over the total number of relevant entities up to rank $k$. The total number of relevant entities for a query is determined by the relevance judgements on a large pool of entities.

\textbf{Mean Average Precision}. MAP provides an overall precision of a retrieval approach across all considered ranks.

\textbf{Normalized Discounted Cumulative Gain}. It takes into account the ranking of entities generated using one of the retrieval approaches and compares it against the ideal ranking in the \emph{ground truth}.
\begin{align*}\label{eq:ndcg}\small{
  nDCG@k = \frac{DCG@{k}}{iDCG@{k}}\;\;\;
  DCG@k = rel_1 + \sum\limits_{i=2}^{k}\frac{rel_i}{log_2{i}}}
\end{align*}
where $DCG@k$ represents the discounted cumulative gain at rank $k$, and $iDCG@k$ is the ideal $DCG@k$ computed from the \emph{ground truth}.

\section{Evaluation and Discussion}\label{sec:evaluation}

In this section we report evaluation results of the two main steps in our approach. We first evaluate the quality of the pre-processing step, i.e., the clustering results for the \emph{x--means} and \emph{spectral} clustering algorithms. Next, we present the findings from our rigorous evaluation of the entity retrieval task.

\subsection{Cluster Accuracy Evaluation}\label{subsec:clustering_eval}
Considering the large number of clusters that are produced in the pre-processing step for a given \textit{type} and \textit{bucket}, evaluating the accuracy and quality of all clusters is infeasible. We randomly select 10 entity types and 10 buckets, resulting in 100 clusters for evaluation, where for each cluster we randomly select a maximum of 10 entities.

To evaluate the \emph{cluster accuracy}, we deploy atomic microtasks modeled such that a worker is presented with sets of 10 entities belonging to a cluster, along with a description of the entity in the form of the entity profile. The task of the worker is to pick the odd entities out (if any). We gather 5 judgments from different workers for each cluster. By enforcing restrictions available on the CrowdFlower platform, and following state of the art task design recommendations, we ensure that we receive judgments from the best workers (workers with high reputation as indicated by CrowdFlower). 

\begin{figure}[ht!]
\centering
	\includegraphics[width=0.7\textwidth]{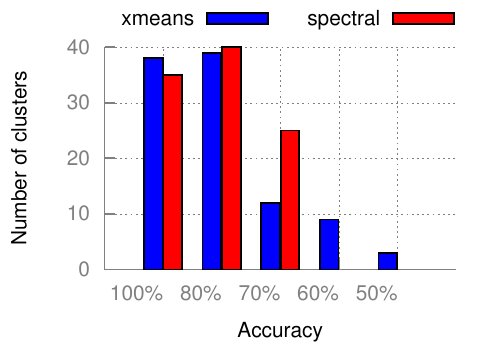}
	\caption{Cluster accuracy for the \emph{spectral} and \emph{x--means} clustering approaches.}
	\label{fig:cluster_accuracy_percentage}
\end{figure}

Figure~\ref{fig:cluster_accuracy_percentage} presents our findings for the evaluation of the clustering process. We note that for \emph{x--means} and \emph{spectral} clustering approaches, nearly 35\% and 38\% of the clusters are judged to be perfect respectively (i.e., the entities within the cluster were all found to belong together). 39\% of the clusters corresponding to \textit{spectral} clustering and 40\% of the clusters corresponding to \textit{x-means}, have an accuracy of 80\%. Considering its multidimensional representation of the entities, \emph{spectral} clustering has higher accuracy and it does not have clusters below 70\% accuracy. 
The lowest accuracy of 70\% for \emph{spectral} clustering implies that in each cluster there were only 3 entities that did not belong to the cluster. The implications of an accurate clustering process become clearer in the next section, where we assess the accuracy of finding relevant entities in the generated entity clusters.

\begin{figure}[ht!]
\centering
    \includegraphics[width=0.7\textwidth]{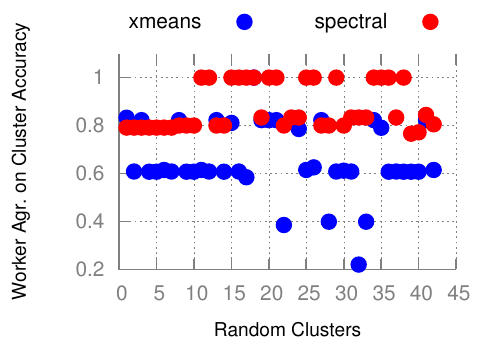}
    \caption{ Worker agreement on cluster accuracy for \emph{spectral} and \emph{x--means} clustering. }
    \label{fig:cluster_agreement}
\end{figure}%

Figure \ref{fig:cluster_agreement} presents the pairwise agreement between workers on the quality of each cluster. In case of the \emph{spectral} clustering, we observe a high inter-worker agreement of 0.75 as per Krippendorf's Alpha.
We observe a moderate inter-worker agreement of 0.6 as per Krippendorf's Alpha on the clusters resulting from \emph{x--means}.

\subsection{Entity Retrieval Evaluation}\label{subsec:retrieval_evaluation}

\begin{figure}[ht!]
\centering
    \includegraphics[width=0.7\textwidth]{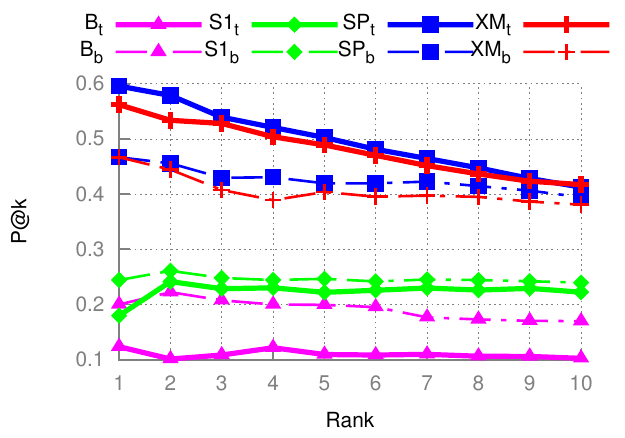}
    \caption{P@k for the different entity retrieval approaches under comparison.}
    \label{fig:precision_k}
\end{figure}

Figure \ref{fig:precision_k} presents a detailed comparison between the $P@k$ for the different methods. The proposed approaches outperform the baseline and state of the art at all ranks. The precision is highest at $P@1=0.6$ whereas for the later ranks it stabilizes at 0.4. In contrast to our approach, the performance of the baseline and the state of the art is more uniform, and is around $P@k=0.25$. The best overall performing approach is the retrieval approach based on spectral clustering $SP$. Table~\ref{tbl:overal_results} shows the details about the performance of the respective approaches as measured for our evaluation metrics.

\begin{figure}[ht!]
\centering
    \includegraphics[width=0.7\textwidth]{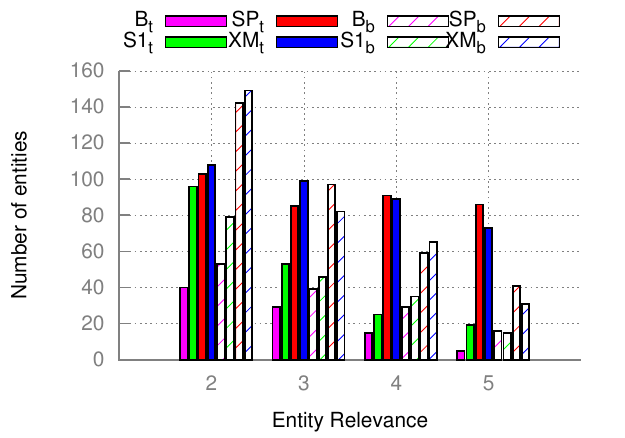}
    \caption{The relevant entity frequency based on their graded relevance (from \textit{2-Slightly Relevant} to \textit{5-Highly Relevant}) for the different methods.}
    \label{fig:relevance_histogram}
\end{figure}

An interesting observation is that for our approaches the best performance is achieved when querying for the field \emph{title}. In the case of the baseline, the best performance is achieved when querying for the field \emph{body} ($B_b$) while the same is inconclusive in case of the state-of-the-art methods (\textbf{$S1_t$} and \textbf{$S1_b$}). We achieve a significantly higher retrieval performance when using the title field. This can be explained by the fact that entities that match a query on their \textit{title} field when compared to those that match a query on their \textit{body} field, have a higher likelihood of being an exact match. 

The high gain in performance through our methods (\textit{SP} and \textit{XM}) stems mainly from the two steps in our approach. The first step expands the result set with relevant entities as shown in Figure~\ref{fig:relevance_histogram}. The figure shows the number of relevant entities corresponding to the different grading scales as described in Section \ref{subsec:clustering_eval}. In all cases we note that our methods find more relevant entities. The second step which re-ranks the expanded result set helps in reducing the number of \emph{`non-relevant'} entities. We find that $S1_t$ has a 14\% decrease of non-relevant entities, whereas $SP_t$ and $XM_t$ depict a 35\% decrease, respectively. In second case where we query the \emph{body} field, the number of \emph{`non-relevant'} entities for $S1_b$ decreases by about 13\%, while $SP_b$ and $XM_b$ depict a 24\% decrease. 

\begin{table}[ht!]\small
\centering
\begin{tabular}{ p{1.2cm} p{1.2cm} p{1.2cm} p{1.2cm} p{1.2cm} p{1.2cm} p{1.2cm} p{1.2cm} p{1.2cm}}
\toprule
& $B_t$ & $B_b$ & $S1_t$ & $S1_b$ & $SP_t$ & $SP_b$ & $XM_t$ & $XM_b$\\
\midrule
P@10 & 0.103 & 0.170 & 0.222 & 0.240 & 0.413 & 0.394 & \textbf{0.417} & 0.381\\
R@10 & 0.052 & 0.089 & 0.112 & 0.118 & 0.206 & \textbf{0.219} & 0.216 & 0.215\\
MAP & 0.110 & 0.191 & 0.224 & 0.246 & \textbf{0.497} & 0.426 & 0.482 & 0.407\\
$Avg(R)$ & 0.031 & 0.058 & 0.063 & 0.074 & 0.132 & \textbf{0.133} & 0.131 & 0.130\\
\bottomrule
\end{tabular}
\caption{\small{Performance of the different entity retrieval approaches. In all cases our approaches are significantly better in terms of P/R ($p<0.05$ measured for \emph{t-test}) compared to \emph{baseline} and \emph{state of the art}. There is no significant difference between $SP$ and $XM$ approaches.}}
\label{tbl:overal_results}
\end{table}

\begin{figure}[ht!]
        \centering
        \includegraphics[width=0.7\textwidth,keepaspectratio]{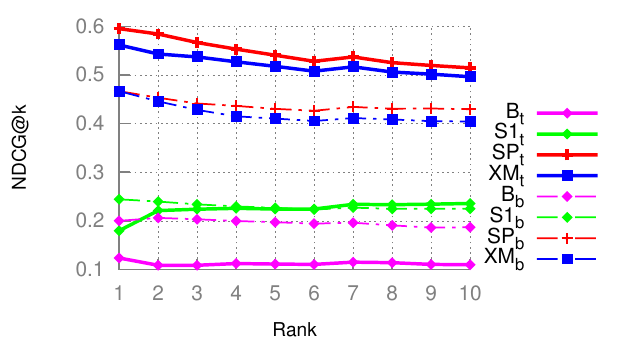}
        \caption{\small{NDCG$@k$ for $B1$, $S1$ and $SP$, $XM$}}
        \label{fig:ndcg}
\end{figure}

We additionally analyze the performance of the entity retrieval approaches through the $NDCG@k$ metric. Figure~\ref{fig:ndcg} shows the NDCG scores. Similar to our findings for $P@k$ presented in Table 1, our approaches perform best for the query field \emph{title} and significantly outperform the approaches under comparison.

\begin{figure}[ht!]
\centering
        \includegraphics[width=0.7\textwidth]{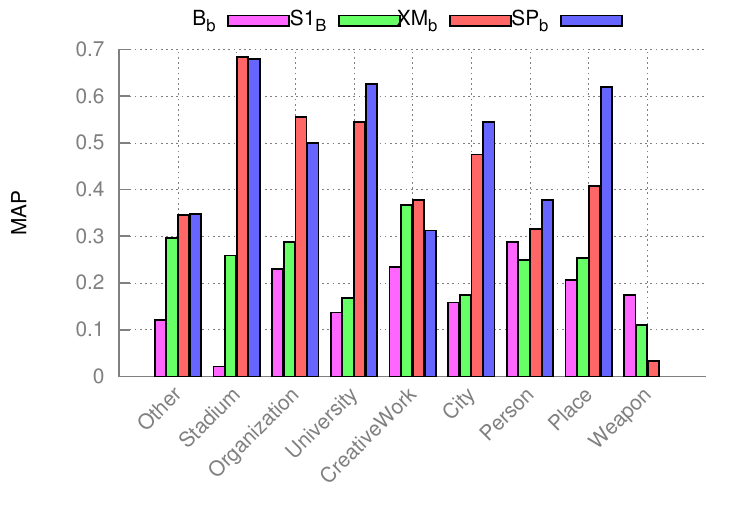}
        \caption{The aggregated MAP for different query types and for the different retrieval approaches (note, we show the results for field \emph{body} where baseline performs best).}
        \label{fig:query_improvement}
\end{figure}

Next, we present observations concerning the different \emph{query types} and the entity result set expansion (see Section~\ref{subsec:entity_similarity}) parameters. In Figure~\ref{fig:query_improvement} we show the improvement we gain in terms of MAP for the different \emph{query types}. We observe that there is quite a variance for the different query types, however, in nearly all cases, the biggest improvement is achieved through the $SP$ approach. Interestingly for the query type \emph{`Creative Work'} the state of the art is nearly as good as the $XM$ approach, whereas in the case of \emph{`Weapon'} the baseline performs best. One possible explanation for this is that in the case of \emph{`Creative Work'} the explicit entity similarity statements are abundant.

\begin{figure}[ht!]
	\centering
        \includegraphics[width=0.7\textwidth]{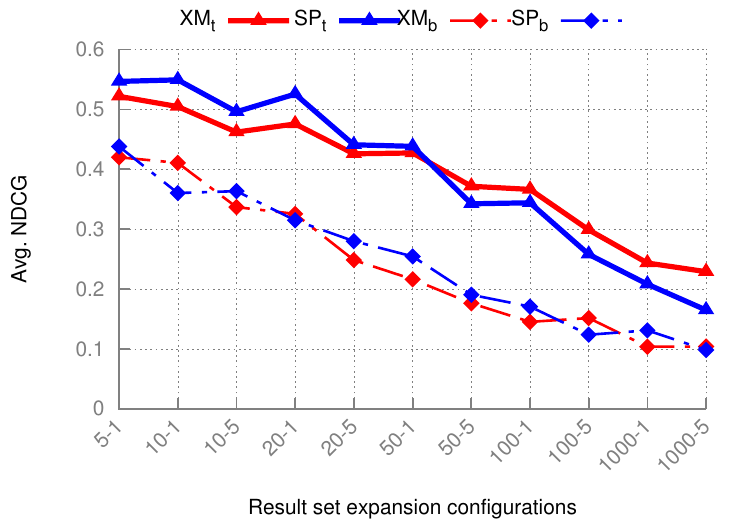}
        \caption{The various configurations for the number of expanded entities for $SP$ and $XM$.}
        \label{fig:result_expansion}
\end{figure}

Addressing the case of optimizing our retrieval approaches, $SP$ and $XM$, we experimentally show the impact that the expansion of the result set has on the measured performance metrics. Here, we show the impact on the \textit{average NDCG} score. Figure~\ref{fig:result_expansion} shows the performance at average NDCG for the varying \emph{cluster size} and \emph{number of entities} added (result set expansion) for every entity in $E_b$. The best performance is achieved for a rather smaller \emph{cluster size} ranging between 5 and 10 entities per cluster. Regarding the number of entities with which the result set is expanded for every $e_b$, the best performance is achieved by expanding with one entity per cluster. The increase in cluster size and number of entities attributes to a decrease in performance.

\subsection{Discussion}\label{subsec:discussion}
\textbf{Scalability. } In the pre-processing stage we introduced the clustering approaches, which first bucket entities together based on the LSH algorithm. This particular step significantly improves the \emph{scalability} of such an offline step. If considering the \emph{x-means} algorithm, under the simplistic assumption that it represents the original \emph{k--means} for which the complexity is $\mathcal{O}(n^{dk+1} log (n))$ (we assume the number of dimensions for the Euclidean space is fixed) for a fixed number of clusters and dimensions. Now, clustering without the bucketing step, we would have around $n=77,485$ entities for clustering with an average of $k=13$ clusters. Hence, $\mathcal{O}(77485^{d\cdot 13+1} log(77485)) > \mathcal{O}(400^{d\cdot 13+1} log(400))$, where after bucketing we have on average $n=400$. Thus, we see a significant decrease in the runtime (while the complexity in theory remains of the same magnitude). For the case of \emph{spectral} clustering this is even more evident, where for the adjacency matrix we consider $n(n-1)/2$ entity pairs, and its singular value decomposition (dependent on the algorithm used) is cubical in terms of big-O notation. 

\textbf{Crowdsourced Evaluation: Precautions.} In order to ensure that we acquire reliable responses from the crowd workers, we take several precautions while designing the tasks for the evaluation of clusters, as well as establishing the ground-truth for the retrieval of entities. We provide clear instructions and examples to avoid misinterpretations in the relevance scoring, leading to a bias in the judgements. We compensate workers with monetary incentives that are proportionate to their contribution. In addition, we use \textit{gold standard questions} as recommended by previous works to curtail malicious activity.

\textbf{Caveats and Limitations.} Considering the optimization of the pre-processing step, the process scales well even for large datasets like the BTC. The retrieval task itself is an online process with no complex approaches and hence the corresponding computational overhead is negligible for the user. We acknowledge the need to re-cluster entities periodically in order to maintain a persistently good entity retrieval performance. However, we believe that this is a relatively minor overhead, when compared to the improvement in performance that it brings about, and given the fact that it is an offline process which can be scaled using parallel infrastructure.

\section{Conclusions and Future Work}\label{sec:discussion}
In this work, we presented an approach to improve the performance of entity retrieval on structured data. Building on existing state of the art methods, we follow an approach consisting of offline preprocessing clustering, and online retrieval, results expansion and reranking. Preprocessing exploits \emph{x--means} and \emph{spectral} clustering algorithms using lexical as well as structural features. The clustering process was carried out on a large set of entities (over 450 million). The evaluation of the clustering process shows that over 80\% of clusters have an accuracy of more than 80\%. As part of the online entity retrieval, for a given a starting result set of entities as retrieved by the baseline approach BM25F we further expand the result set with relevant entities. Additionally, we propose an entity ranking model that takes into account the query type affinity. Finally, we carry out an extensive evaluation of the retrieval process using the SemSearch and the BTC12 datasets. The results show that our methods outperform the baseline and state of the art approaches. In terms of standard IR metrics, our method in combination with one of the clustering approaches, e.g. $SP_t$ improves over $S1_t$ with $\Delta P@10 = +0.19$, $\Delta MAP= +0.273$ and $\Delta R@10 = +0.1$. 

\paragraph{\textbf{Acknowledgements:}} This work is partly funded by the EU under FP7 project DURAARK (grant no. 600908) and ERC Advanced Grant ALEXANDRIA (grant no. 339233).

\end{document}